\documentclass[12pt]{article}

\catcode`\@=11
\global\arraycolsep=2pt
\usepackage{amsbsy,amssymb,latexsym,amsfonts,amsmath}
\usepackage{graphicx,color}

\begin{document}
\begin{flushright}
\parbox{4.2cm}
{RUP-18-24}
\end{flushright}

\vspace*{0.7cm}

\begin{center}
{ \Large Gravity Dual for Very Special Conformal Field Theories 

in type IIB Supergravity}
\vspace*{1.5cm}\\
{Yu Nakayama}
\end{center}
\vspace*{1.0cm}
\begin{center}

Department of Physics, Rikkyo University, Toshima, Tokyo 171-8501, Japan

\vspace{3.8cm}
\end{center}

\begin{abstract}
We study holographic dual descriptions of very special conformal field theories with the T(2) symmetry. After constructing solutions in effective five dimensional Einstein gravity coupled with massive two-form fields, we uplift them to the ten dimensional type IIB supergravity, via a consistent truncation ansatz, to derive new analytical solutions in string theory. From the Kaluza-Klein ansatz in terms of the internal Sasaki-Einstein space, we obtain their field theory interpretation with concrete realizations in a large class of holographic $\mathcal{N}=1$ supersymmetric conformal field theories. Null compactification of these theories yields holographic dual descriptions of non-relativistic critical systems with translational invariance but without rotational invariance such as the ones induced from a constant electromagnetic field.

\end{abstract}

\thispagestyle{empty} 

\setcounter{page}{0}

\newpage


%-----------------------------------------

%\pacs{}
%\keywords{Suggested keywords}%Use shokeys class option if keyword
%\maketitle

%%%%%%%%%%%%%%%%%%%%%%%%%%%%%%%%%%%%%%%%%%%%%%%%%%%%
\section{Introduction}
In high energy physics, the Poincar\'e symmetry is often regarded as a fundamental assumption, but it is theoretically possible that it is merely an accidental low energy symmetry. One way to break the Lorentz symmetry while preserving the translational symmetry is to  choose a particular direction in space-time. Certainly in our universe, there is a special ``time direction" which emerges from the evolution of the universe, resulting in the breaking of the Lorentz symmetry down to a spatial rotation. Mathematically, however, it is more interesting to choose a particular null direction to break the Lorentz symmetry, leading to more non-trivial residual symmetries than the mere spatial rotation. This was suggested by Cohen and Glashow sometime ago \cite{Cohen:2006ky}\cite{Cohen:2006ir} and dubbed ``very special relativity".

To address the question of the emergent Lorentz symmetry, we recall that our modern understanding of effective field theories is based on the idea of renormalization group, and it is instructive to see how quantum field theories with the very special relativity behaves under the renormalization group flow. In order to study the renormalization group flow, it is crucial to understand the structure of  renormalization group fixed points with the very special relativity. In \cite{Nakayama:2017eof}, we have classified all the conformal extension of the very special relativity as possible candidates of the renormalization group fixed points that are consistent with the very special relativity. Such theories were named ``very special conformal field theories".

A remarkable feature of the construction of very special conformal field theories is that if we start with the four-dimensional Poincar\'e invariant conformal field theories, the  deformation that preserves the very special conformal symmetry has the Poincar\'e conformal dimension of five, which means (i) it is an irrelevant deformation from the point of view of the Poincar\'e conformal scaling and (ii) it may appear to be ``non-renormalizable" from the naive power-counting. Therefore, while it is easy to write down the effective action that preserves the very special conformal symmetry at the tree level, the study of their quantum nature is difficult by design.

In this paper, we offer an alternative view on the strongly interacting very special conformal field theories possibly realized at the ultraviolet fixed point by using the holography. The holographic dual or gravity dual descriptions of strongly coupled conformal field theories are successful not only with Poincar\'e invariance but also without Poincar\'e invariance. In our previous paper \cite{Nakayama:2017eof}, we have found that one of such examples, i.e. holographic dual description of the Schr\"odinger invariant conformal field theories \cite{Nishida:2007pj}\cite{Son:2008ye}\cite{Balasubramanian:2008dm}\cite{Adams:2008wt}\cite{Maldacena:2008wh}\cite{Herzog:2008wg}, has many common features with the holographic dual description of a particular kind of very special conformal field theories with the E(2) symmetry. See also earlier works \cite{Bergman:2000cw}\cite{Alishahiha:2003ru} on the holographic descriptions of these theories with the same symmetry.
The main focus of this paper is to study the least symmetric case of the T(2) invariant very special conformal field theories and their gravity duals which have not been addressed before.  As we will see, null compactification of these theories yields holographic dual descriptions of non-relativistic critical systems with translational invariance but without rotational invariance and may have potential applications in condensed matter physics.

The organization of the paper is follows. In section 2, we review the basic facts about the very special conformal field theories. In section 3, we construct the holographic duals in the effective five dimensional Einstein gravity coupled with two-form matter fields. In section 4, we uplift the solutions to ten dimensional type IIB supergravity and obtain new analytical solutions. In section 5, we give the field theory interpretation of our holographic dual solutions. In section 6, we conclude with further discussions.

\section{Very Special Conformal Field Theories}
Very special conformal symmetry is a subgroup of the Poincar\'e conformal symmetry of $SO(4,2)$ that preserves a particular null direction and include at least one special conformal transformation. For our purpose, let us choose this null direction to be $x^-$, where $x^{\pm} = \frac{1}{\sqrt{2}}(x^0\pm x^1)$ with the metric convention $ds^2 = -2dx^+dx^- + (dx^i)^2$. Here $i=2,3$.

Based on the classification without the conformal symmetry \cite{Cohen:2006ky}\cite{Cohen:2006ir}, in  \cite{Nakayama:2017eof} we  argued that there are four interesting classes of very special conformal symmetry. The maximal case is SIM(2) very special conformal algebra spanned by $\{P_+,P_-,P_i,J_{+i},J_{+-},J_{ij},\tilde{D},K_+\}$, whose commutation relation can be inferred from the original Poincar\'e conformal group with the conventional notation of $P_\mu$ being translation, $J_{\mu\nu}$ being Lorentz transformation, $K_\mu$ being special conformal transformation, and $D$ being dilatation (while $\tilde{D} = D + J_{+-})$. If we abandon $J_{ij}$, we obtain HOM(2) very special conformal algebra, and if we abandon $J_{+-}$, we obtain E(2) very special conformal algebra. Finally, if we abandon $J_{+-}$ and $J_{ij}$ we obtain T(2) very special conformal algebra, which will be the main focus of this paper. Note that the names only refer to the commutation relations of the Lorentz part of the algebra. For completeness the commutation relations of T(2) very special algebra are summarized in Table 1.

\begin{table}[htb] 
  \begin{tabular}{|c|cccccc|} \hline
     & $P_+$ & $P_-$ & $P_i$  & $J_{+i}$ & $K_+$ & $\tilde{D}$ \\ \hline 
$P_+$ & $0$ & $0$ & $0$   & $0$  & $0$ & $0$ \\ 
$P_-$ & $0$ & $0$ & $0$   & $P_i$ & $-\tilde{D}$ & $2P_-$ \\ 
$P_i$ & $0$ & $0$ & $0$   & $P_+$  &$J_{+i}$ & $P_i$ \\ 
$J_{+i}$ & $0$ & $-P_i$ & $-P_+$  & $0$ & $0$ & $-J_{+i}$ \\ 
$K_+$ & $0$ & $\tilde{D}$ & $-J_{+i}$ & $0$ & $0$ & $-2K_+$ \\ 
$\tilde{D}$ & $0$ & $-2P_-$ & $-P_i$   &$J_{+i}$ & $2K_+$ & $0$ \\ \hline
  \end{tabular} \label{tab1}
\caption{The commutation relation of the T(2) very special conformal algebra.}
\end{table}

%The idea of Eebedding Formalism is to realize (a subgroup of) the conformal algebra linearly in higher dimensional null cone. Let us consider the $(D+2)$ dimensional space $R^{D,2}$ with the embedding metric
%\begin{align}
%ds^2 = \eta_{\mu\nu} dX^\mu dX^\nu - dX^{+'} dX^{-'} \ . 
%\end{align}
%Since we reserve $+$ and $-$ for the $D$-dimensional part of the light-cone direction, we used $+'$ and $-'$ for the extra coordinate here.
%On this space, the $D+2$ dimensioanl Lorentz transformation $J_{MN}$ acts linearly. We identify them as $SO(2,D)$ conformal algebra as
%\begin{align}
%J_{\mu\nu} &= J_{\mu\nu} \cr
%J_{\mu+'} & = P_\mu \cr
%J_{\mu-'} &= K_\mu \cr
%J_{+'-'} & = D 
%\end{align}

Very special conformal field theories possess a local energy-momentum tensor that satisfies the following conditions
\begin{align}
\partial_+T_{+}^{\ +} + \partial_-T_{+}^{\ -} + \partial_i T_+^{\ i} &= 0 \cr
\partial_+T_{-}^{\ -} + \partial_- T_{-}^{\ -} + \partial_i T_-^{\ i} &= 0 \cr
\partial_+{T_j}^{\ +} + \partial_-T_{j}^{\ -} + \partial_i T_{j}^{\ i} &= 0 \cr
T_{i}^{\ -} + T_{+}^{\ i} & = 0 \cr
2T_{-}^{\ -} + T_{i}^{\ i} &=0 \  .
\end{align}
 so that one can construct the local current associated with the dilatation $\tilde{D}$ and the special conformal transformation $K_+$.

\section{Holographic realizations}
The main purpose of this paper is to construct holographic descriptions of the T(2) very special conformal field theories while the corresponding holographic models for E(2) very special conformal field theories were constructed in \cite{Nakayama:2017eof}.\footnote{The cases with SIM(2) and HOM(2) very special conformal field theories are much more challenging because we have to abandon either unitarity or locality. See some ideas to construct holographic duals in \cite{Nakayama:2018fib}.}
 In the field theory side, one way to construct very special conformal field theories with T(2) symmetry is to start with the Poincar\'e invariant conformal field theories and deform them by using the dimension five primary operator in the anti-symmetric tensor representation of the Poincar\'e group.  
\begin{align}
S = S_0 + \int d^4x O^{-x} ,
\end{align}
where $S_0$ is the action with Poincar\'e conformal invariance, and $O^{-x} = -O^{x-}$ transforms as an anti-symmetry tensor and it has $\Delta = 5$ under the Poincar\'e dilatation $D$.

To mimic the construction in the gravity side, we begin with the five dimensional Einstein-Hilbert action with negative cosmological constant, which supports the AdS space, coupled with a massive two-form tensor field, which is dual to the dimension five anti-symmetric tensor operator.  The explicit action is 
\begin{align}
S_B = \int d^5x \sqrt{g} \left(\frac{1}{2}R - \Lambda - \frac{1}{6} H_{MNL} H^{MNL} - \frac{m^2}{2} B_{MN} B^{MN} \right) \label{effective}
\end{align}
where $B_{MN} = -B_{NM}$ and $H_{MNL} = \partial_M B_{NL} + \partial_N B_{LM} + \partial_L B_{MN}$. We will set $\Lambda = -6$ and $m^2 = 9$,  where we recall that with this cosmological constant the mass and the Poincar\'e conformal dimension $\Delta$ of the two-form field are related as $m^2 = (\Delta-2)^2$.

We find the solution of the equations motion that have the symmetry corresponding to the $T(2)$ very special conformal symmetry. 
The metric is given by
\begin{align}
ds^2 = -c \frac{dx^{-2}}{z^4} + \frac{dz^2-2dx^+dx^{-} + (dx^2)^2 + (dx^3)^2}{z^2} \ ,
\end{align}
where it is invariant under the dilatation $\tilde{D}$ 
\begin{align}
x^- \to \lambda^2 x^- \ , \ \ x^i \to \lambda x^i \ , \ \ z \to \lambda z \ , \ \ x^+ \to x^+ 
\end{align}
as well as the very special conformal transformation $K_+$
\begin{align}
x^- \to \frac{x^-}{1+ax^{-}} \ , \ \ x^i \to \frac{x^i}{1+a x^-} \ , \cr
 z \to \frac{z}{1+a x^{-}} \ , \ \ x^+ \to x^+ - \frac{a}{2}\frac{x_i^2 + z^2}{1+ax^{-}} \ . 
\end{align}
Note that this metric has the larger symmetry than the T(2) symmetry because the rotation in $ij$ directions is not broken: it is invariant under the E(2) symmetry. One can argue that whenever one wants to preserve the T(2) symmetry in the five dimensional metric, it must possess the enhanced E(2) symmetry.

In order to break the isometry further down to the T(2) very special conformal symmetry, we need a non-trivial matter configuration. In our case, the massive two-form tensor has the condensation
\begin{align}
B = \frac{1}{2} B_{MN} dx^{M} dx^N = b \frac{dx^{-}dx^2}{z^3} \ ,
\end{align}
which is invariant only under the isometry of the T(2) very special conformal symmetry: $J_{ij}$ rotation is obviously broken by choosing a particular $x^2$ direction in the two-form. We can readily compute the matter energy-momentum tensor and it has only component in $--$. Then the $--$ component of the Einstein equation demands $c=6b^2$. Note that $c$ can be changed by the rescaling of $dx^-$ and $dx^+$ unless it vanishes, so its  magnitude is not physically significant. We also note that although $B$ breaks the $ij$ rotation, the energy-momentum tensor constructed out of $B$ does not break the $ij$ rotation so that the Einstein equation is consistent with the above metric ansatz.

In $d=5$ dimensions, we may use an alternative way to introduce a massive two-form field from the topological coupling. Let us consider a complex two-form field $q_2$ with the action
\begin{align}
S_q = \int \left(\frac{1}{2}R *1 - \Lambda *1 -\frac{i}{3}\bar{q}_2 \wedge dq_2 + \frac{i}{3}q_2 \wedge d\bar{q}_2 -2q_2\wedge *\bar{q}_2 \right) \ , 
\end{align}
where $*$ is the five dimensional Hodge dual operator. The equation of motion of $q_2$ gives the self-dual equation
\begin{align}
dq_2 = 3i * q_2 
\end{align}
which describes a propagating (real) massive two-form with $m^2 = 9$. To see this, we  may substitute $q_2 = k_2 + i l_2$ with real two-form fields $k_2$ and $l_2$. The equations of motion tell that $l_2 = \frac{1}{3} * dk_2$ and $k_2$ (or accordingly $l_2$ as well) satisfies the massive two-form equation with $m^2 = 9$. The Einstein equation is 
\begin{align}
R_{MN} = -4 g_{MN} +2(q_{M}^{L} \bar{q}_{N L} + q_{N}^{L} \bar{q}_{M L} - \frac{1}{12}g_{MN} q_{L K} \bar{q}^{LK}) 
\end{align}

We find the following solutions with the symmetry corresponding to the T(2) very special conformal symmetry: the metric is given by
\begin{align}
ds^2 = -c \frac{dx^{-2}}{z^4} + \frac{dz^2-2dx^+dx^{-} + (dx^2)^2 + (dx^3)}{z^2}
\end{align}
and the  massive two-form has the condensation
\begin{align}
q_2 = q (\frac{dx^{-}dx^2}{z^3} + i \frac{dx^{-}dx^3}{z^3})\ .
\end{align}
The $--$ component of the Einstein equation demands $c = \frac{4}{3} q^2$.

\section{Uplifting to type IIB supergravity}
In the previous section, we constructed the gravity dual for the T(2) very special conformal field theories in the bottom up approach. Here we would like to uplift them to the ten dimensional type IIB supergravity, allowing us to interpret them from the string theory viewpoint as we will discuss in the next section.

Our starting point is the $AdS_5 \times SE_5$ flux compactification of type IIB supergravity 
\begin{align}
ds_{10}^2 &= ds^2(AdS_5) + ds^2(SE_5) \cr
F_{(5)} &= 4 \mathrm{vol}(SE_5) + 4\mathrm{vol}(AdS_5) \cr
H_{(3)}&= F_{(3)}=0 \ 
\end{align}
with a constant axion-dilaton field.
Here the Sasaki-Einstein space $SE_5$ has a preferred $U(1)$ isometry called Reeb Killing vector and the metric has the local decomposition of
\begin{align}
ds^2(SE_5) = ds^2(KE_4) + \eta^2 ,
\end{align}
where $\eta = d\psi + \mathcal{A}$ is the (contact) one-form dual to the Reeb Killing vector.
On the Sasaki-Einstein space, there is an $SU(2)$ structure $(\eta,J,\Omega)$ which satisfies
\begin{align}
J\wedge \Omega &= 0 \cr
\Omega \wedge \bar{\Omega} &= 2 J \wedge J = 4 \mathrm{vol}(KE_4) \cr
*_4 J &= J \cr
*_4 \Omega &= \Omega \cr
d\eta &= 2J \cr
d\Omega &= 3i \eta \wedge \Omega \ .
\end{align}
Here $*_4$ is the Hodge dual on the (locally) Kahler Einstein base $KE_4$. The Kahler form $J$ and $(2,0)$ form $\Omega$ are induced from those of the Kahler-Einstein base $KE_4$.
The holographic interpretation of these gravity solutions have been studied in the literature and they are given by  $\mathcal{N}=1$ superconformal field theories realized as large $N$ quiver gauge theories \cite{Klebanov:1998hh}\cite{Benvenuti:2004dy}.

With these ingredients, our Kaluza-Klein ansatz is to deform the five-form flux as
\begin{align}
ds^2_{10} &= ds^2(E_5) + ds^2(KE_4) + \eta^2 \cr
F_{(5)} &= 4 \mathrm{vol}(SE_5) + 4\mathrm{vol}(E_5) + (* q_2 \wedge \Omega + q_2 \wedge \Omega \wedge \eta + c.c.)  \cr
F_{(3)} &= 0 \cr
H_{(3)} &= 0 \cr
C_{(0)} &= 0 \cr
\Phi &= \phi \ , \label{KK}
\end{align}
where $*$ is the Hodge dual with respect to the five dimensional metric $ds^2(E_5)$.
For the complete consistent truncation, we have to keep the other scalar degrees of freedom such as the breathing mode, but for our particular solution, we are not going to excite any scalar degrees of freedom as we will argue so that this ansatz will become consistent.

As discussed in \cite{Liu:2010sa}\cite{Gauntlett:2010vu}, the Kaluza-Klein reduction gives us the five dimensional Einstein gravity coupled with massive (complex) two-form field $q_2$ that has the topological mass term with $m^2 = 9$, which we have already discussed in the previous section. 
The non-trivial claim is that the Kaluza-Klein reduction is a consistent truncation and the five dimensional solution
\begin{align}
ds^2 &= -c \frac{dx^{-2}}{z^4} + \frac{dz^2-2dx^+dx^{-} + (dx^2)^2 + (dx^3)}{z^2} \cr
q_2 &= q (\frac{dx^{-}dx^2}{z^3} + i \frac{dx^{-}dx^3}{z^3})\ .
\end{align}
substituted into \eqref{KK} solves the ten-dimensional type IIB supergravity equations of motions when $c = \frac{4}{3} q^2$. Most of the relevant computation has been done in the literature \cite{Liu:2010sa}\cite{Gauntlett:2010vu}, but since it is a relatively simple solution, one can directly check that the ten dimensional equations of motion hold. 

Explicitly, the five form equations of motion gives
\begin{align}
dq_2 = 3i * q_2 
\end{align}
and it is readily solved by our ansatz since it is the same equation as the self-dual equation of the complex two-form field. The self-dual condition of the five form was already encoded in our Kaluza-Klein ansatz. Then the $MN$ (i.e. five dimensional) component of the ten dimensional Einstein equation gives
\begin{align}
R_{MN} = -4 g_{MN} +2(q_{M}^{L} \bar{q}_{N L} + q_{N}^{L} \bar{q}_{M L} - \frac{1}{12}g_{MN} q_{L K} \bar{q}^{LK})  \label{einstein}
\end{align}
and this is exactly the same equations that we have already solved in the effective five dimensional gravity. The crucial point here is that the other equations of motion are not affected because in those equations we have to contract $q_2$ to make a scalar, but since $q_2$ has only a null component, the scalar contraction always gives zero. Another point is that the Einstein equation \eqref{einstein} holds not only in average but pointwise in the internal space due to the special geometric feature of our Kaluza-Klein ansatz.  

%One thing to be noted is that the extra scalar degrees of freedom such as the Warp factor needed for the generic consistent trunction are not excited because the massive two-form condensation contains one null direction, and we need to contract the two-form to obtain the scalar backreaction, but the null contraction gives zero. Similarly, the only backreaction to the metric should be in the $--$ component as is indeed the case in  our solution.

\section{Field theory interpretation}
A comparison of  the Kaluza-Klein ansatz with the linearized harmonic analysis on the flux compactification of the Sasaki-Einstein space with the AdS/CFT dictionary allows us to interpret the holographic solutions we constructed from the dual field theory perspective. As we have already mentioned two-form fields with  the mass $m^2=9$  in the gravity side correspond to dimension five anti-symmetric tensor operators in the dual field theories. More explicitly, in the Sasaki-Einstein compactification, it is given by 
\begin{align}
O = \sum_I c_I \mathrm{Tr} F_I^{x-}(\lambda^I \lambda^I + cc) \ ,
\end{align}
where $\lambda^I$ are gauginos and $I$ runs over the gauge groups of the quiver gauge theory. See e.g. \cite{Ceresole:1999zs} for the case of $SE_5 = T_{1,1}$ and \cite{Gunaydin:1984fk}\cite{Kim:1985ez} for the case of $SE_5=S_5$.
In the general flux compactification of the Sasaki-Einstein space, there is one particular choice of $c_I$  over different gauge groups which have a protected scaling dimension so that one can use it as our dimension five anti-symmetric tensor deformations.
In the case of $\mathcal{N}=4$ super Yang-Mills theory, it is in the $10_c$ rep of SU(4) R-symmetry from the choice of the four different fermions, and all of them can be used. 

The operator $O$ is charged under the R-symmetry as we can also see from the Kaluza-Klein reduction (since $\Omega$ appearing in the ansatz transforms under the Lie derivative with respect to the Reeb Killing vector), and it breaks the $R$-symmetry once added. Note also that the operator breaks (at least part of) the supersymmetry because the protected operator  $ \mathrm{Tr} F_{\dot{\alpha}\dot{\beta}} \lambda \lambda$ and its conjugate   $\mathrm{Tr}F_{{\alpha}{\beta}} \bar{\lambda} \bar{\lambda}$ preserves the different supersymmetry. Even this is the case, the deformation is an exactly marginal deformation with respect to the very special conformal symmetry as demonstrated in the holographic description since the deformation parameter $c$ is arbitrary.

\section{Discussions}
In this paper, we constructed the gravity dual for T(2) invariant very special conformal field theories. In particular, we have presented new exact solutions in type IIB supergravity with the isometry corresponding to the T(2) very special conformal symmetry. Such top-down constructions give precise field theory interpretation of our dual descriptions in terms of explicit $\mathcal{N}=1$ supersymmetric conformal field theories.

Our  field theory interpretation reveals that we can construct very special conformal field theories from the Poincar\'e conformal field theory by introducing dimension five operator such as the gauge field strength multiplied by gaugino bilinear. The similar operator can be found in the standard model of particle physics. For example, in QED one can readily construct the dimension five anti-symmetric tensor operator such as $F^{x-} \bar{\Psi}_e\Psi_e$, where $\Psi_e$ is the electron field, and in this way one may imagine the Lorentz violating interaction with the T(2) very special conformal symmetry may exist in the (ultraviolet completion of the) nature.

Beside the particle physics application, we would like to point out that the gravity dual description of T(2) symmetric very special conformal field theories have potential applications in condensed matter physics. Suppose we compactify the null direction $x^+$, and regard the Kaluza-Klein momentum as a particle number. Then we obtain non-relativistic conformal field theories with $z=2$ scaling. Here $x^-$ direction is identified with the non-relativistic time, and $J_{+i}$ is regarded as the Galilean boost symmetry. Furthermore $K_+$ is regarded as the non-relativistic special conformal transformation. Now the difference between our (compactified) T(2) invariant non-relativistic critical systems and the ones studied in the literature (e.g. the Sch\"odinger conformal field theories) is that we do not have the spatial rotation as a symmetry. Therefore, our theory is suitable to describe the system without rotational symmetry (but with the translational symmetry). Such critical systems are ubiquitous: one may realize them by introducing background electric field or magnetic field along a particular spatial direction. It would be interesting to investigate their properties such as thermal properties by constructing corresponding blackhole solutions in our setup.

\section*{Acknowledgements}
This work is in part supported by JSPS KAKENHI Grant Number 17K14301. The preliminary results of this paper were presented at KIAS-YITP 2017 "Strings, Gravity and Cosmology". The author would like to thank J.~Gauntlett and N.~Kim for fruitful comments and lively discussions there.

%%%%%%%%%%%%%%%%%%%%%%%%%%%%%%%%%%%%%%%%%%%%%%%%%%%%%%%%%%%%%%%%%%%%%%%%%%%%%%%%


\begin{thebibliography}{99}
%\cite{Cohen:2006ky}
\bibitem{Cohen:2006ky} 
  A.~G.~Cohen and S.~L.~Glashow,
  %``Very special relativity,''
  Phys.\ Rev.\ Lett.\  {\bf 97}, 021601 (2006)
  doi:10.1103/PhysRevLett.97.021601
  [hep-ph/0601236].
  %%CITATION = doi:10.1103/PhysRevLett.97.021601;%%
  %186 citations counted in INSPIRE as of 09 Jul 2017

%\cite{Cohen:2006ir}
\bibitem{Cohen:2006ir} 
  A.~G.~Cohen and S.~L.~Glashow,
  %``A Lorentz-Violating Origin of Neutrino Mass?,''
  hep-ph/0605036.
  %%CITATION = HEP-PH/0605036;%%
  %53 citations counted in INSPIRE as of 09 Jul 2017


%\cite{Nakayama:2017eof}
\bibitem{Nakayama:2017eof} 
  Y.~Nakayama,
  %``Very special conformal field theories and their holographic duals,''
  Phys.\ Rev.\ D {\bf 97}, no. 6, 065003 (2018)
  doi:10.1103/PhysRevD.97.065003
  [arXiv:1707.05423 [hep-th]].
  %%CITATION = doi:10.1103/PhysRevD.97.065003;%%
  %2 citations counted in INSPIRE as of 10 Jul 2018



%\cite{Nishida:2007pj}
\bibitem{Nishida:2007pj} 
  Y.~Nishida and D.~T.~Son,
  %``Nonrelativistic conformal field theories,''
  Phys.\ Rev.\ D {\bf 76}, 086004 (2007)
  doi:10.1103/PhysRevD.76.086004
  [arXiv:0706.3746 [hep-th]].
  %%CITATION = doi:10.1103/PhysRevD.76.086004;%%
  %195 citations counted in INSPIRE as of 09 Jul 2017



%\cite{Son:2008ye}
\bibitem{Son:2008ye} 
  D.~T.~Son,
  %``Toward an AdS/cold atoms correspondence: A Geometric realization of the Schrodinger symmetry,''
  Phys.\ Rev.\ D {\bf 78}, 046003 (2008)
  doi:10.1103/PhysRevD.78.046003
  [arXiv:0804.3972 [hep-th]].
  %%CITATION = doi:10.1103/PhysRevD.78.046003;%%
  %609 citations counted in INSPIRE as of 09 Jul 2017

%\cite{Balasubramanian:2008dm}
\bibitem{Balasubramanian:2008dm} 
  K.~Balasubramanian and J.~McGreevy,
  %``Gravity duals for non-relativistic CFTs,''
  Phys.\ Rev.\ Lett.\  {\bf 101}, 061601 (2008)
  doi:10.1103/PhysRevLett.101.061601
  [arXiv:0804.4053 [hep-th]].
  %%CITATION = doi:10.1103/PhysRevLett.101.061601;%%
  %572 citations counted in INSPIRE as of 09 Jul 2017



%\cite{Adams:2008wt}
\bibitem{Adams:2008wt} 
  A.~Adams, K.~Balasubramanian and J.~McGreevy,
  %``Hot Spacetimes for Cold Atoms,''
  JHEP {\bf 0811}, 059 (2008)
  doi:10.1088/1126-6708/2008/11/059
  [arXiv:0807.1111 [hep-th]].
  %%CITATION = doi:10.1088/1126-6708/2008/11/059;%%
  %272 citations counted in INSPIRE as of 09 Jul 2017

%\cite{Maldacena:2008wh}
\bibitem{Maldacena:2008wh} 
  J.~Maldacena, D.~Martelli and Y.~Tachikawa,
  %``Comments on string theory backgrounds with non-relativistic conformal symmetry,''
  JHEP {\bf 0810}, 072 (2008)
  doi:10.1088/1126-6708/2008/10/072
  [arXiv:0807.1100 [hep-th]].
  %%CITATION = doi:10.1088/1126-6708/2008/10/072;%%
  %312 citations counted in INSPIRE as of 09 Jul 2017

%\cite{Herzog:2008wg}
\bibitem{Herzog:2008wg} 
  C.~P.~Herzog, M.~Rangamani and S.~F.~Ross,
  %``Heating up Galilean holography,''
  JHEP {\bf 0811}, 080 (2008)
  doi:10.1088/1126-6708/2008/11/080
  [arXiv:0807.1099 [hep-th]].
  %%CITATION = doi:10.1088/1126-6708/2008/11/080;%%
  %270 citations counted in INSPIRE as of 09 Jul 2017

%\cite{Bergman:2000cw}
\bibitem{Bergman:2000cw} 
  A.~Bergman and O.~J.~Ganor,
  %``Dipoles, twists and noncommutative gauge theory,''
  JHEP {\bf 0010}, 018 (2000)
  doi:10.1088/1126-6708/2000/10/018
  [hep-th/0008030].
  %%CITATION = doi:10.1088/1126-6708/2000/10/018;%%
  %95 citations counted in INSPIRE as of 11 Jul 2017

%\cite{Alishahiha:2003ru}
\bibitem{Alishahiha:2003ru} 
  M.~Alishahiha and O.~J.~Ganor,
  %``Twisted backgrounds, PP waves and nonlocal field theories,''
  JHEP {\bf 0303}, 006 (2003)
  doi:10.1088/1126-6708/2003/03/006
  [hep-th/0301080].
  %%CITATION = doi:10.1088/1126-6708/2003/03/006;%%
  %69 citations counted in INSPIRE as of 11 Jul 2017

%\cite{Nakayama:2018fib}
\bibitem{Nakayama:2018fib} 
  Y.~Nakayama,
  %``Local field theory construction of Very Special Conformal Symmetry,''
  arXiv:1802.06489 [hep-th].
  %%CITATION = ARXIV:1802.06489;%%

%\cite{Liu:2010sa}
\bibitem{Liu:2010sa} 
  J.~T.~Liu, P.~Szepietowski and Z.~Zhao,
  %``Consistent massive truncations of IIB supergravity on Sasaki-Einstein manifolds,''
  Phys.\ Rev.\ D {\bf 81}, 124028 (2010)
  doi:10.1103/PhysRevD.81.124028
  [arXiv:1003.5374 [hep-th]].
  %%CITATION = doi:10.1103/PhysRevD.81.124028;%%
  %68 citations counted in INSPIRE as of 11 Jul 2018


%\cite{Gauntlett:2010vu}
\bibitem{Gauntlett:2010vu} 
  J.~P.~Gauntlett and O.~Varela,
  %``Universal Kaluza-Klein reductions of type IIB to N=4 supergravity in five dimensions,''
  JHEP {\bf 1006}, 081 (2010)
  doi:10.1007/JHEP06(2010)081
  [arXiv:1003.5642 [hep-th]].
  %%CITATION = doi:10.1007/JHEP06(2010)081;%%
  %81 citations counted in INSPIRE as of 11 Jul 2018

%\cite{Klebanov:1998hh}
\bibitem{Klebanov:1998hh} 
  I.~R.~Klebanov and E.~Witten,
  %``Superconformal field theory on three-branes at a Calabi-Yau singularity,''
  Nucl.\ Phys.\ B {\bf 536}, 199 (1998)
  doi:10.1016/S0550-3213(98)00654-3
  [hep-th/9807080].
  %%CITATION = doi:10.1016/S0550-3213(98)00654-3;%%
  %996 citations counted in INSPIRE as of 19 Jul 2018

%\cite{Benvenuti:2004dy}
\bibitem{Benvenuti:2004dy} 
  S.~Benvenuti, S.~Franco, A.~Hanany, D.~Martelli and J.~Sparks,
  %``An Infinite family of superconformal quiver gauge theories with Sasaki-Einstein duals,''
  JHEP {\bf 0506}, 064 (2005)
  doi:10.1088/1126-6708/2005/06/064
  [hep-th/0411264].
  %%CITATION = doi:10.1088/1126-6708/2005/06/064;%%
  %241 citations counted in INSPIRE as of 11 Jul 2018

%\cite{Ceresole:1999zs}
\bibitem{Ceresole:1999zs} 
  A.~Ceresole, G.~Dall'Agata, R.~D'Auria and S.~Ferrara,
  %``Spectrum of type IIB supergravity on AdS(5) x T**11: Predictions on N=1 SCFT's,''
  Phys.\ Rev.\ D {\bf 61}, 066001 (2000)
  doi:10.1103/PhysRevD.61.066001
  [hep-th/9905226].
  %%CITATION = doi:10.1103/PhysRevD.61.066001;%%
  %184 citations counted in INSPIRE as of 11 Jul 2018


%\cite{Gunaydin:1984fk}
\bibitem{Gunaydin:1984fk} 
  M.~Gunaydin and N.~Marcus,
  %``The Spectrum of the s**5 Compactification of the Chiral N=2, D=10 Supergravity and the Unitary Supermultiplets of U(2, 2/4),''
  Class.\ Quant.\ Grav.\  {\bf 2}, L11 (1985).
  doi:10.1088/0264-9381/2/2/001
  %%CITATION = doi:10.1088/0264-9381/2/2/001;%%
  %319 citations counted in INSPIRE as of 12 Jul 2018

%\cite{Kim:1985ez}
\bibitem{Kim:1985ez} 
  H.~J.~Kim, L.~J.~Romans and P.~van Nieuwenhuizen,
  %``The Mass Spectrum of Chiral N=2 D=10 Supergravity on S**5,''
  Phys.\ Rev.\ D {\bf 32}, 389 (1985).
  doi:10.1103/PhysRevD.32.389
  %%CITATION = doi:10.1103/PhysRevD.32.389;%%
  %508 citations counted in INSPIRE as of 12 Jul 2018


\end{thebibliography}
\end{document}